\documentclass[conference]{IEEEtran}
\IEEEoverridecommandlockouts
\usepackage{cite}
\usepackage{amsmath,amssymb,amsfonts}
\usepackage{algorithmic}
\usepackage{graphicx}
\usepackage{textcomp}
\usepackage{xcolor}
\usepackage{subfig}
\usepackage{units}
\usepackage{mathtools}

\usepackage[hidelinks]{hyperref}
\usepackage{pgfplots}
\usepackage{tikz}
\usepackage[T1]{fontenc}
\usepackage[utf8]{inputenc}
\usepackage{pgfplots}
\usepackage{grffile}
\pgfplotsset{compat=newest}
\usetikzlibrary{plotmarks}
\usetikzlibrary{arrows.meta}
\usepgfplotslibrary{patchplots}

\usepackage{enumitem}

\usepackage{url}
\usepackage{adjustbox}
\usepackage[labelsep=period]{caption}
\usepackage{booktabs,array}
\usepackage{multirow}
\usepackage{balance}
\usepackage{colortbl}%

\definecolor{mycolor1}{rgb}{0.00000,0.44700,0.74100}%
\definecolor{mycolor2}{rgb}{0.45000,0.79000,0.52000}%
\definecolor{mycolor3}{rgb}{0.93000,0.69000,0.13000}%
\definecolor{mycolor4}{rgb}{0.85000,0.33000,0.10000}%

\begin{document}
	
	\title{DimRad: A Radar-Based Perception System \\for Prosthetic Leg Barrier Traversing}
	
	\author{\IEEEauthorblockN{Fady Aziz, Bassam Elmakhzangy, Christophe Maufroy, Urs Schneider, Marco F.~Huber
			\thanks{*The first two authors contributed equally to this work.}}
		\IEEEauthorblockA{ Fraunhofer Institute for Manufacturing Engineering and Automation IPA, Stuttgart, Germany}
	}
	
	\maketitle
	
	\begin{abstract}
		Lower extremity amputees face challenges in natural locomotion, which is partially compensated using powered assistive systems, e.g., micro-processor controlled prosthetic legs. In this paper, a radar-based perception system is proposed to assist prosthetic legs for autonomous obstacle traversing, focusing on multiple-step staircases. The presented perception system is composed of a radar module operating with a multiple-input-multiple-output (MIMO) configuration to localize consecutive stair corners.  An inertial measurement unit (IMU) is integrated for coordinates correction due to the angular dispositioning that occurs because of the knee angular motion. The captured information from both sensors is used for staircase dimensioning (depth and height). A shallow neural network (NN) is proposed to model the error due to the hardware limitations and enhance the dimension estimation accuracy ($\approx$ \unit[1]{cm}). The algorithm is implemented on a microcontroller subsystem of the radar kit to qualify the perception system for embedded integration in powered prosthetic legs. 
	\end{abstract}
	
\section{Introduction} \label{intro}
	\par Many people undergo amputations in their lower limbs every year due to accidents and medical conditions such as vascular diseases and complications associated with diabetes and cancer \cite{amputationCauses}. Approximately 185,000 amputation operations are carried out each year in the USA, with an estimation that this number will be doubled by 2050 \cite{amputationUSA}. To partially restore the lost mobility, patients who suffer from above-the-knee amputations can rely on a powered prosthetic leg to emulate a semi-natural gait movement with minimal walking fatigue \cite{roboticAnkle}. The main challenges in developing a natural motion of such prostheses are the mechanical design, efficient motor control of the joints, and motion planning \cite{joints}. 
	
	\par Environmental sensing is a main feature in prosthetic leg to support autonomous obstacle traversing. For this purpose, a scanning sensor of real-time capturing capability with respect to the maximum achievable walking velocity is required. Moreover, information regarding the prosthetic limb locomotion is often required to identify the current walking gait phase (i.e., swing or stance), which is commonly gathered using an inertial measurement unit (IMU) \cite{surv1}. In recent years, vision-based modules, such as depth cameras and laser scanners, have been extensively studied for the task of classifying and dimensioning surrounding obstacles \cite{surv2,surv3}. These sensors are usually of high refresh rate and perception resolution  \cite[Fig.~3]{surv1}. Thus, they satisfy both the real-time constraint and detection accuracy requirement. However, such systems depend on 3D point cloud acquisitions, and they require a high computational budget to segment and dimension the surrounding objects\cite{depthcamera}. Additionally, the vision-based sensors are affected significantly by different lighting and weather conditions, e.g., complete darkness or underclothes. To overcome these environmental limitations, non-visionary solutions should be considered.
	
	\par  Radar sensors have been selected for human-interaction applications due to their capability of real-time processing with respect to the human motion and their measuring capability under any environmental conditions~\cite{8835652,8455723}.  Moreover, they have been presented as feasible sensors for complex obstacles detection, especially stairs \cite{laribi1_11, radarSherif}. These studies are based on a frequency modulation continuous wave (FMCW) radar operating with a single-input-single-output (SISO) configuration, which can only provide 1D-range perception. To overcome this limitation, \cite{laribi1_11} proposed using multi-path information to estimate the height of a curbstone or a single-step based on the non-line-of-sight reflection coming from the edge of a curbstone due to the knife-edge diffraction phenomenon. However, this algorithm is limited to single-step scenarios. In \cite{radarSherif}, the authors formulated a 2D-scan of the sagittal-plane based on an external mechanical motion of the radar. This resulted in a high angular resolution perception at the expense of real-time processing and computational complexity. Similarly, \cite{radarBernard} relied on the prosthetic-knee angular motion to formulate a 2D-scan of the sagittal-plane, which needs continuous monitoring of the walking gait. This was achieved by relying on IMU and a machine learning-based technique to identify the swing phase, i.e., tibia motion for 2D-scan formulation. Thus, the detection accuracy is dependent on the walking gait realization. To the best of our knowledge, this publication is the only available integration of radar as a perception system in powered prosthetic legs. 
	
	\par In this study, a radar-based perception system (DimRad) is utilized to assist microprocessor-controlled prosthetic legs in the stair-climbing process. In contrast to \cite{radarSherif,radarBernard}, our module is based on integrating a multiple-input-multiple-output (MIMO) radar module with an IMU sensor. The MIMO operation is based on electronic beamforming, and this will mitigate the need for any external mechanical motion to construct 2D-scans of the sagittal-plane. Thus, the perception capability of our system is independent of the walking gait. However, this will come at the expense of lower angular resolution, which will affect the estimated staircase depth and height accuracy. To this end, we propose a shallow neural network (NN) for enhancing the estimated dimensions using the given hardware obligations. To further reduce the required computational complexity, the introduced dimensioning algorithm is based on localizing stair corners instead of intensive evaluation of the 2D-scans.

	\begin{figure*}[!t]
		\includegraphics[width=\textwidth]{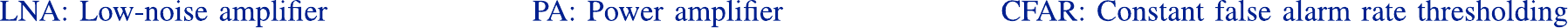}\vspace{-3mm}
		\subfloat[Radar/IMU aligned on acrylic plate.]
		{\label{fig:acrylic_plate}{\includegraphics{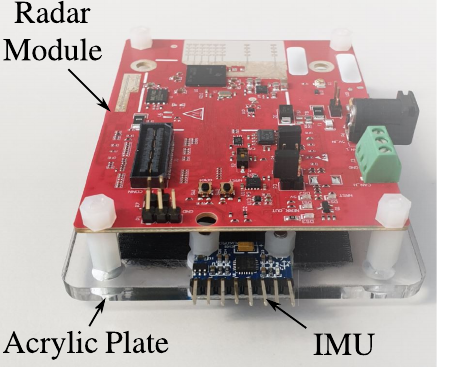}}}
		\hfill
		\subfloat[An overview of the utilized embedded controllers.]
		{\label{fig:cont}{\includegraphics{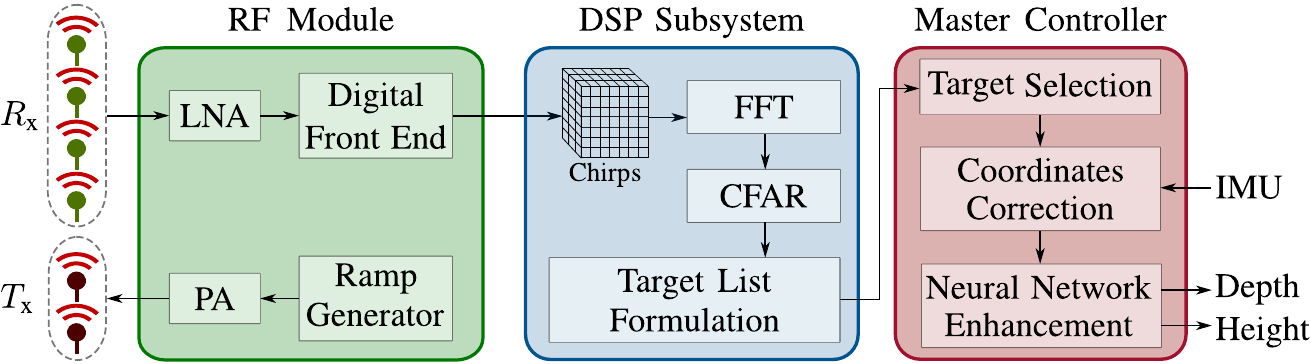}}}
		\caption{The proposed perception system with detailed information of the embedded subsystems.}
			\vspace{-4mm}
		\label{fig:setup}
	\end{figure*}
	
	\section{Perception System Description} \label{percsystem}
	\par The radar module and the IMU are mounted parallel to each other on an acrylic plate, as shown in Fig.~\ref{fig:acrylic_plate}. The radar module selected for the perception system is a TI-mmwave IWR1642 chip operating at \unit[77-81]{GHz} \cite{Rub1}. As depicted in Fig.~\ref{fig:cont}, the selected radar chip is featured with three subsystems: the radio frequency (RF) module, digital signal processing (DSP) subsystem, and master controller, respectively. The RF module contains the analog circuits and microcontroller, which sets the RF parameters and controls transmission/reception. The DSP subsystem contains a microcontroller for implementing the signal processing chain and defining a target list with coordinates in the 2D space. The master controller is where the stair dimensioning task is implemented based on the estimated angular positioning by the IMU and the formulated target list by the DSP. Accordingly, the dimensioning task is carried out through these steps:
	\begin{itemize}
		\item Filtering the target list for localizing consecutive corners.
		\item Coordinates correction based on the IMU estimations.
		\item Stair depth and height estimation.
		\item Accuracy enhancement based on a shallow NN.
	\end{itemize}
	
	\par  Operating in a millimeter-wave spectrum reduces the size of the antennas and allows the radar module to be compact and suitable for integration in a prosthetic limb. Additionally, since all the algorithms are implemented on the microcontrollers of the radar chip, the system can be integrated into the control printed circuit board (PCB) of the prosthesis. 
	
	\vspace{-2mm}\subsection{RF Module}~\label{rf_sec}\vspace{-4mm}
	\par An FMCW radar has been selected for the presented study due to its capability for estimating the mapped range and velocity of the detected target \cite{lipa1990fmcw}. The operation principle of the FMCW radar is based on transmitting periodic chirps modulated at a certain carrier frequency ($f_o$) within a specific transmission bandwidth ($B$). Each chirp is transmitted with an adjustable period ($T_{ch}$). The time delay between the transmitted and the received signals can be used to estimate the target range from the radar ($r$). Both the maximum detectable range ($r_{max}$) and the minimum distinguishable range between two targets ($r_{res}$) can be adjusted as follows:
	
	\begin{equation}
	r_{res}=\frac{c}{2B},  ~ r_{max} = \frac{cN_S}{2B}
	\label{eqn:2}
	\end{equation}
	where $N_S$ is the number of samples per chirp, and $c$ is the speed of light in space. The target velocity can be measured by estimating the frequency shift induced due to the Doppler effect through monitoring $N_P$ consecutive received chirps. The velocity resolution $v_{res}$ (minimum distinguishable velocity between two targets) is parameterized based on:
	
	\begin{equation}
	v_{res} = \frac{c}{2f_oT_{ch}N_P}
	\label{eq:vers}
	\end{equation}
	
	\par To formulate real-time 2D-scans of the sagittal-plane, an FMCW radar with single-input-multiple-output (SIMO) configuration can be used for electronic beam steering \cite{geibig2016compact}. The SIMO operation is based on deploying multiple receiving antennas, which are assembled to be equally-spaced with half of the transmission wavelength ($\lambda$) to avoid grating lobes~\cite{shoykhetbrod2012design}. As described in \cite{milligan2005modern}, the \unit[3]{dB} angular resolution ($\alpha_{res}$) can be defined as:  
	\begin{equation}
	\alpha_{res} = \frac{1.78}{N_A}
	\label{eq:thres}	
	\end{equation}
	where $N_A$ is the number of receiving antennas. From this relation, we can conclude that the higher the number of receiving antennas, the better is the angular resolution. Thus, the technique presented here is based on a MIMO configuration, that deploys two transmitting antennas to simulate the effect of a doubled number of receiving antennas \cite{mimo_beamforming}. The utilized transmission protocol simulates the effect of 8 receiving antennas using 6 physical antennas (2 transmitters and 4 receivers) instead of 9 antennas (1 transmitter and 8 receivers). The radar module is parameterized as shown in Table~\ref{tab:param}.  
		\begin{table}[!t]	
			\setlength\arrayrulewidth{0.05pt}
			\def\arraystretch{1.15}
			\Large
			\centering
			\caption{Utilized MIMO radar module parametrization.  \label{tab:param}\vspace{-2mm}}
			\resizebox{\columnwidth}{!}{
				\begin{tabular}{l r l r}
					\toprule
					\multicolumn{2}{c}{Radar Parametrization} & \multicolumn{2}{c}{Range \& Velocity Attributes} \\
					\midrule
					Carrier frequency ($f_o$)  & \unit[77]{GHz} & {} & {}\\
					Transmitting-Receiving antennas  & 2-4 & Maximum range ($r_{max}$) & \unit[6]{m}\\
					Bandwidth ($B$)  & \unit[3.6]{GHz} & Range resolution ($r_{res}$)& \unit[4.16]{cm}\\
					Chirp duration ($T_{ch}$)   & \unit[64]{$\mu s$} & Velocity resolution ($v_{res}$) & \unit[3.8]{m/s}\\
					Samples per chirp ($N_S$)   & 144 & {} & {}\\
					Chirps per measurement ($N_P$)   & 8 & {} & {}\\
					\bottomrule
				\end{tabular}
			}
			\vspace{-5mm}
		\end{table}
	%	Given that an FMCW radar is used as the base module, a range-AoA map can be generated for each estimated velocity as shown in Fig.~\ref{fig:heatmap}.
	
	\subsection{DSP Subsystem}~\label{scan}
	
	\vspace{-4mm}
	
	\begin{figure*}[t]
		\centering
		\includegraphics[width=\textwidth]{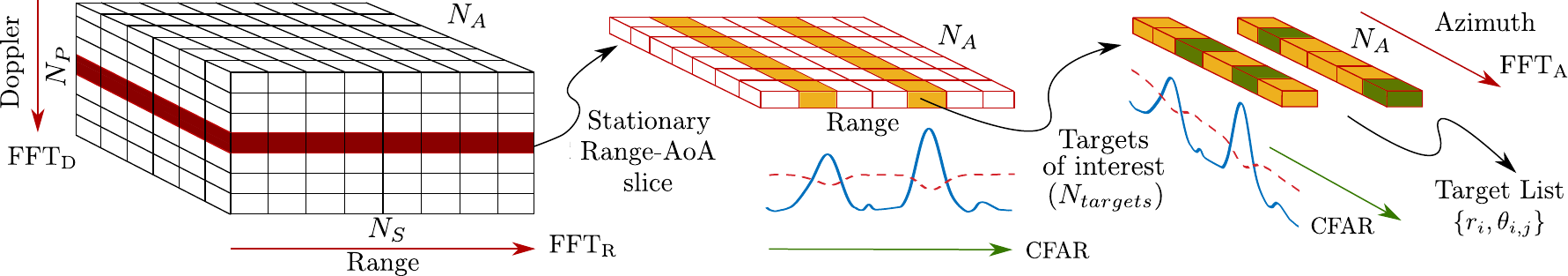}
		%	\vspace{-0.5 cm}
		\caption{Illustration of the proposed low-complexity target list formulation algorithm.\label{fig:3dfft}}	
		\vspace{-0.5cm}
	\end{figure*}
	
	%	\par The electronic beam-forming offers real-time 2D-scans for the sagittal plane without any further requirement for external mechanical motion as the scanning techniques presented in \cite{radarBernard, radarSherif}. In the first study, a motor is utilized to rotate the radar for acquiring range profiles from different angles and achieve a 2D-scan of the sagittal plane. While, the second scanning technique is relying on the tibia motion for generating the 2D-scan, which can be affected by different walking styles and velocities. Moreover, both aforementioned techniques can only offer single scan within one walking gait cycle. On the other hand, the electronic beam-forming enables acquisition of multiple 2D-scans with low latency. However, acquiring with low latency is achieved at the expense of the angular resolution, which affects the estimation accuracy of the height dimension. The technique presented in \cite{8902510,denoise} proposes a generative adversarial networks (GANs) super-resolution to avoid the high-resolution hardware requirement. However, this technique is not suitable for either the walking real-time constraint or the embedded integration intention. Accordingly, the presented algorithm in this study is not based on processing through the full 2D-scan and does not encounter any image processing. Thus, the proposed methodology in Sec.\ref{method} incorporates the utility of MIMO radar and embedded integration in terms of design, power and walking motion constraints.     
	
	\par The main task of the DSP subsystem is generating a target list of the detected, stationary targets (expected to be stair corners). The MIMO configuration is utilized for estimating the range, angle of arrival (AoA), and velocity simultaneously for each target. The chirp acquisitions are distributed in a 3D matrix with ($N_S \times N_P \times N_A$) dimensions as shown in Fig.~\ref{fig:3dfft}. Then, an N-point FFT is applied on each matrix dimension, revealing mapped data voxels (range-AoA-velocity), where each bin step is evaluated based on the feature resolution defined by \eqref{eqn:2}, \eqref{eq:vers} and \eqref{eq:thres} respectively. A 2D-FFT is first applied on the $N_S \times N_P$ plane of the acquired 3D-chirp matrix resulting in a range-Doppler map \cite{lipa1990fmcw}. Since the radar acquisitions are captured while walking, the stationary stairs will be detected by the radar with a velocity component equivalent to the walking velocity of the prosthetic limb ($-v_{host}$). The maximum velocity of the prosthetic leg should be equivalent to the normal feet velocity within the walking gait $\approx$ \unit[3]{m/s} \cite{boulic1990global}. Accordingly, the radar is parameterized with $v_{res} >|v_{host}|$. Thus, the stationary range-AoA slice ($0<|v_{host}|<v_{res}$), shown in Fig.~\ref{fig:3dfft}, should include any stationary target in the radar field of view. Henceforth, any non-stationary target will be detected by the radar with a velocity component ($>v_{res}$) will be filtered out, e.g., other people climbing down the stairs. After extracting the range-AoA slice, the outline of the proposed methodology for generating the target list, described in Fig.~\ref{fig:3dfft}, proceeds as follows:
	
	\begin{itemize} %[label=$\bullet$,wide = 0pt]
		\item Accumulating the range profiles over the $N_A$ dimension. 
		\item To detect stationary targets, a constant false alarm rate (CFAR) algorithm \cite{cfar} is applied on the accumulated range profile. The targets ($N_{targets}$) exceeding the adaptive CFAR threshold are singled out.
		\item To estimate the AoA profiles, an $\textrm{FFT}_{\textrm{A}}$ is applied on the $N_A$ dimension through the $N_{targets}$ bins. Thus, an exhaustive FFT operation for all $N_{S}$ bins is avoided.
		\item An additional CFAR operation is applied on the resultant AoA profiles.
	\end{itemize}

	\begin{figure*}[t]
		\subfloat[Overview of the prothetic limb with the stairs.\label{fig:mdiag}]{\includegraphics[width=0.65\textwidth]{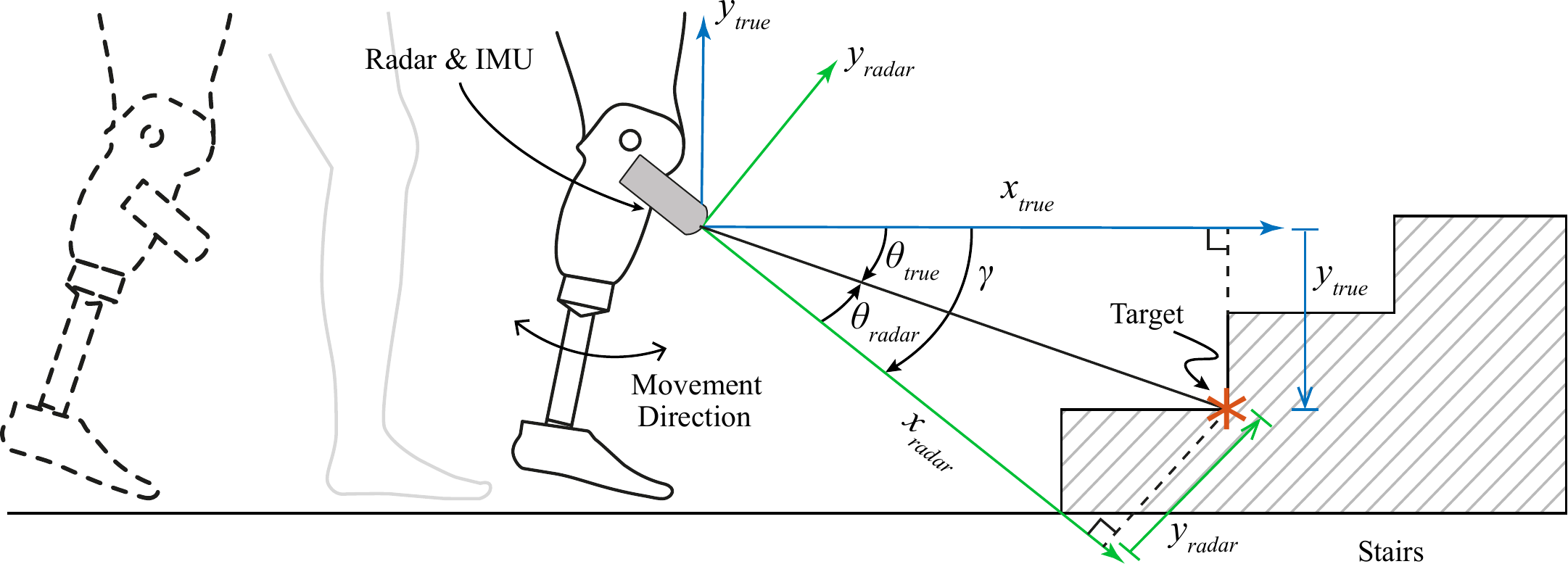}}
		\hfill
		\subfloat[Initial dimensioning procedure.\label{fig:dim}]{\includegraphics[width=0.3\textwidth]{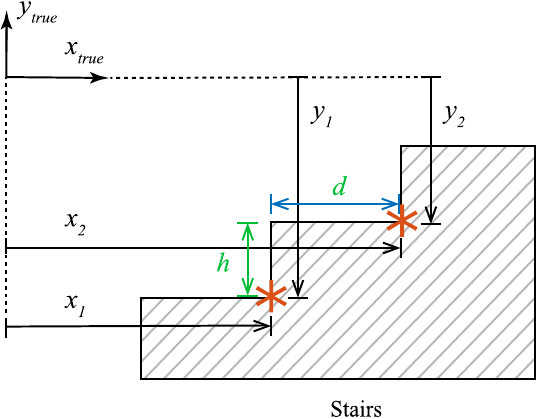}}
		\caption{A schematic of the experimental setup used for data acquisitions with all relevant symbols.\label{fig:schematic}}
		\vspace{-4mm}
	\end{figure*}
	
	\par Finally, a target list $\left\{r_i,\theta_{i,j}\right\}$ is formulated, where $r_i$ represents the detected range and $\theta_{i,j}$ is the detected $j^{th}$ AoA for the $i^{th}$ target. Since the radar scans are acquired in the vertical plane, the target list will include multiple AoAs for each target range. Henceforth, the list is conveyed to the master controller for identifying the expected targets to be stair corners. Afterwards, the stair dimensioning process is accomplished by the master controller through fusing between the joint information from the DSP subsystem and the IMU.

	\subsection{Master Controller (Coordinates Correction)} \label{Sec:mastCC}
	\par During the walking gait cycle, the radar tilting is influenced by the prosthetic angular locomotion. Thus, the estimated coordinates  $\left\{r_i,\theta_{i,j}\right\}$ for each target cannot reflect true positions. To accommodate for this, the IMU is used to measure the inclination of the perception system with respect to the floor ($\gamma$). The walking gait scenario, shown in Fig.~\ref{fig:mdiag}, describes all the detected relevant angles when a staircase is available in the detection field of view. 
	For corners localization, a conversion from the polar coordinates $\left\{r_i,\theta_{i,j}\right\}$ to Cartesian coordinates $\left\{x_i,y_{i,j}\right\}$ is required. Thus, the aforementioned tilting ($\gamma$) will be utilized to calculate the true target polar and Cartesian coordinates as follows:
	
	\begin{subequations}\label{first:correction}
		\begin{equation}
		\theta^{t}_{i,j} = \gamma \pm \theta_{i,j}  \label{eq:th_correction}
		\end{equation}
		\vspace{-2mm}
		\begin{equation}
		x^{t}_{i} = r_i \cdot \cos(\theta^{t}_{i,j}) \hspace{10mm} y^{t}_{i,j} = r_i \cdot \sin(\theta^{t}_{i,j}) \label{eq:xy_correction}
		\end{equation}
	\end{subequations}
	where $\left\{x^{t}_i,y^{t}_{i,j}\right\}$ and $\left\{r^{t}_{i},\theta^{t}_{i,j}\right\}$ represent the true Cartesian and polar coordinates, respectively. Afterwards, the dimensions of stairs can be estimated by identifying two targets from the target list that are suspected to be two consecutive corners. Based on the studies presented in \cite{knott2012radar}, stair corners and sharp edges should yield the highest reflection power. To filter out these corners or edges for each step in the list, a commonly used standard for depth (\unit[22-35]{cm}) and height (\unit[10-22]{cm}) is used as a ground-truth reference \cite[p.~2]{roys2001serious}. These standards are formulated to avoid stair injuries and for a more comfortable climbing experience. 
	
	\par Eventually, the coordinates of the two consecutive corners are used to calculate the depth $d$ and the height $h$ of steps, as shown in Fig.~\ref{fig:dim}. This step is applied to all targets in the list until both $d$ and $h$ fulfill the aforementioned standards. Thus, this condition will ensure that only consecutive corners are figured out in the target list. Since the proposed approach considers uniform dimensions, therefore the estimated depth $d$ and the height $h$ are generalized for all steps in a single staircase acquisition. However, this assumption can be violated in irregular terrain scenarios (outdoor), which forces the first step to differ from other steps in the same staircase by $\approx$ \unit[1]{cm}. This is still considered as a safe error margin.
	
	\subsection{Master Controller (Accuracy Enhancement)}\label{sec:nn}
	\par The estimated corner coordinates are subject to random and systematic errors, which degrade the accuracy of stairs dimensioning. In the proposed system, the main sources of error are the radar hardware noise and the walking motion of the prosthetic limb. The radar hardware noise induces an error following a standard Normal distribution. However, the error induced due to the non-linear walking motion of the prosthetic limb is usually of an unknown distribution. This non-linearity is attributed to the change of the radar inclination angle and the measured ranges/AoAs, as shown in Fig.~\ref{fig:mdiag}.
	
	\par Tracking, filtering, and learning-based techniques are frequently used to mitigate such dimensioning errors and hardware limitations \cite{radarSherif,8902510,denoise}. Such techniques require high computational complexity that may consume additional power and affect real-time constraint. Accordingly, they are unsuitable for compact-size mobile robotic modules, e.g., prostheses, because most of the available power is preserved for motion functionality. 
	
	%	Moreover, these techniques are based on feeding previous observations together with the present state to get a better estimate of the current measurement. 
	%	
	%	However, the perception system in a real use case will collect non-equidistant observations as in many inclinations no targets are observed in the field of view. 
	%	
	%	Thus, such techniques are not a good option for enhancing the measurements as they enforce time causality. 
	
	%Consequently, the proposed NN architecture is shallow to consume little computational resources (time and power). 
	
	%	 At the end of Sec.~\ref{Sec:dimEst}, the two consecutive corners obeying the aforementioned standards are singled out from the target list of a single radar acquisition.
	
	\par
\begin{figure}  
		\centering
		\includegraphics[width=0.9\columnwidth]{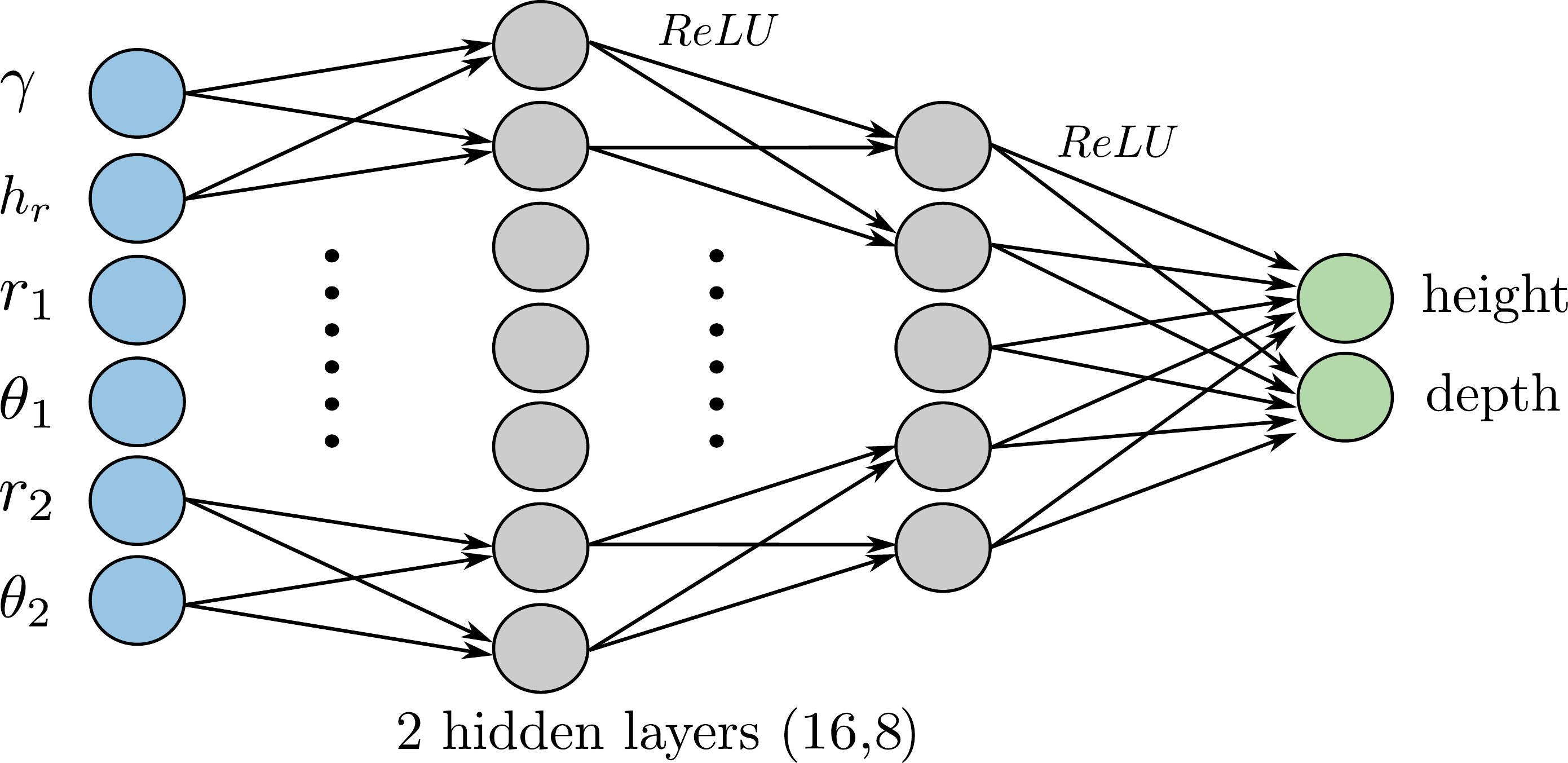}
		\caption{The shallow NN taking radar and detected targets positions as input to estimate the stairs height and depth.}
		\label{fig:nn_layers}
		\vspace{-5mm}
	\end{figure}

 For these reasons, a shallow NN is used in this paper as a suitable trade-off remedy for both power and real-time requirements. The main functionality of the NN is to derive the relation between the current erroneous corner coordinates and the demanded stair dimensions. As shown in Fig.~\ref{fig:nn_layers}, a 3-layered NN with (16,8) neurons is trained to estimate a given depth and height. The corrected polar coordinates of two consecutive corners are fed to the NN together with the IMU inclination $\gamma$ and the current radar height $h_r$. Since the radar and the IMU are mounted on the prosthesis with an angle of $-20^\circ$, the tibia would be inclined by $\gamma+20^{\circ}$. Therefore, the current radar height $h_r$ in centimeters can be calculated as a function of the initial height $h_i$ according to: 
	\vspace{-0.5mm}
	\begin{equation}
	h_r = h_i \cdot \cos(\gamma+20^\circ) \label{eq:hr}
	\vspace{-0.5mm}
	\end{equation}
	
	Namely, the network will take as an input $r_1,\theta_1,r_2,\theta_2,h_r,\gamma$, representing the corners polar coordinates, radar height and inclination, respectively.
	
%	\begin{equation}
%	h_r = h_i \cdot \cos(\gamma+20^\circ) \label{eq:hr}
%	\end{equation}
	
	\vspace{-0.25mm}
	\section{Experimental Setup} \label{expSetup}
	\par Based on \cite{boulic1990global}, the lower leg length is $\approx 25\%$ of the human height. In this study, test subjects were chosen with heights \unit[155-195]{cm}. Accordingly, the described perception system in Sec.~\ref{percsystem} is mounted to the tibia of a prosthetic leg below the knee at a height of \unit[40-50]{cm} from the ground and is tilted by \unit[20]{$^\circ$}. This tilting is adopted to adjust the angle between the radar and the stairs to the boresight (axis of the maximum gain), resulting in a higher detection capability.
	
	\par To generate the training data set, the perception system is mounted to the tibia of a person below the knee while walking towards the stairs. Each staircase acquisition is collected within a period of $\approx \unit[5]{s}$ and a distance to the stairs between \unit[4]{m} and \unit[0.5]{m}. For each staircase, the radar is mounted on 8 training and 2 testing subjects with variable heights. The perception system is parameterized with a sampling rate of \unit[10]{Hz}. Each sample includes a captured inclination angle ($\gamma$) by the IMU and the corresponding polar coordinates of any detected target in the radar field of view. The targets assumed to be stair corners are singled out according to the stair selection algorithm discussed in Sec.~\ref{Sec:mastCC}. Thus, the collection process is independent of any walking style or specific knee flexing angle.  
	
	\par
\begin{figure}[t]
		\centering
		\includegraphics[width=\columnwidth]{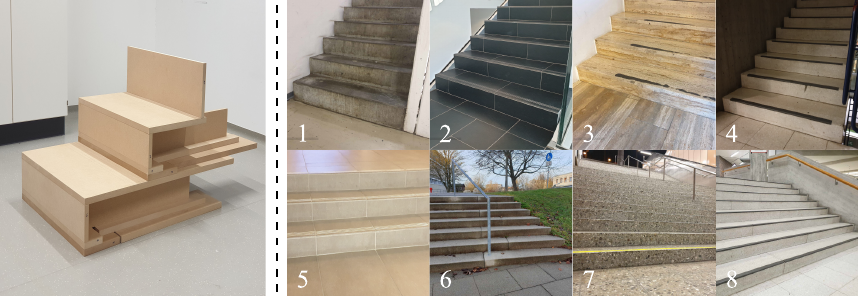}
		\caption{The lab-constructed stairs on the left is used in training set to realize different dimension combinations. The test set on the right cover different dimensions \& materials including: concrete, wood, and ceramic.}
		\vspace{-5mm}
		\label{fig:stairs_cases}
	\end{figure}

 Multiple staircases (45 training and 8 testing scenarios) were included in indoor and outdoor environments, following the common design standards~\cite[p.~2]{roys2001serious}. The used dimension combinations are often repeated, especially in indoor environments. Therefore, to ensure a more generalized performance of the NN, wooden stairs with different dimension combinations were constructed as shown in Fig.~\ref{fig:stairs_cases}. The built-up staircases were selected to cover dimension combinations of \unit[26-38]{cm} for depth and \unit[10-18]{cm} for height that can be adjusted in increments of \unit[2]{cm} on each dimension, resulting in a total of 35 staircase scenarios. Furthermore, 18 additional environments were selected and split into ten training and eight testing environments as shown in Fig.~\ref{fig:stairs_cases}. The NN is trained for 50 epochs until convergence using Adam optimizer.
	
%	Additionally, 18 more environments were collected and splitted into 10 training and 8 testing as shown in Fig.~\ref{fig:stairs_cases}. 
	
	\section{Results and Discussion} \label{results}
	%	\begin{figure}[t]
	%		\centering
	%		\includegraphics[width=\columnwidth]{lab_stairs.pdf}
	%		\caption{Lab-constructed wooden stairs to realize different dimension combinations.}
	%		\label{fig:mdfStairs}
	%	\end{figure}
	%	\begin{figure}[t]
	%		\centering
	%		\includegraphics[width=\columnwidth]{all_stairs.pdf}
	%		\caption{Examples of stair-cases included in the test dataset with materials including: concrete, wood and ceramic.}
	%		\vspace{-5mm}
	%		\label{fig:stairs_cases}
	%	\end{figure}
	
	%Multiple stair-cases composed of different materials were selected for testing in indoor and outdoor scenarios as shown in Fig.~\ref{fig:stairs_cases}. The selected test stair-cases are unseen by the network during training. 
	
	\par To emphasize the importance of the NN-enhancement, we compare the dimensioning accuracy achieved using the initial approach presented in Sec.~\ref{Sec:mastCC} with the NN approach presented in Sec.~\ref{sec:nn}. Moreover, it is important to point out that accuracy comparison against state-of-the-art methods \cite{radarBernard,radarSherif} is not within the scope of this work, as this would require building their entire perception system for such an evaluation. However, they reported an average accuracy of $\approx$ \unit[1]{cm} in both depth and height. The mean absolute error (MAE), the root mean squared error, and the standard deviations ($\sigma$) were used as metrics for evaluating the dimensioning approaches. 
	
%	 as we must realize their whole perception system for such evaluation.
	
	\par
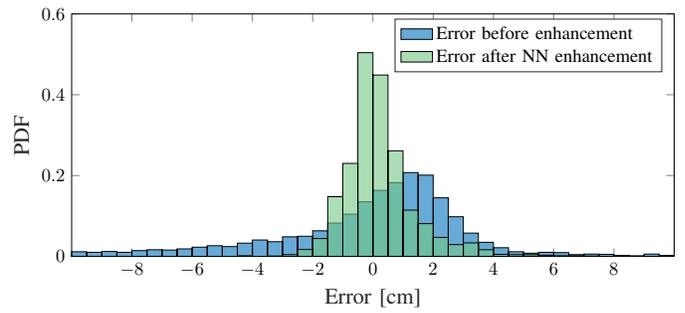
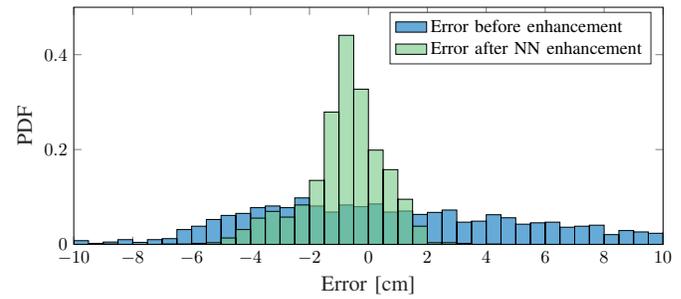
\begin{figure}[h]
		\vspace{-5mm}
		\captionsetup[subfigure]{oneside,margin={1.cm,0cm}}
		\centering
		\hspace{-2mm}\subfloat[Error distribution in stairs depth estimation.\label{fig:deviations_depth}]{\resizebox{\columnwidth}{!}{\begin{tikzpicture}
\begin{axis}[%
width=4.52in,
height=1.82in,
at={(0in,0in)},
scale only axis,
xmin=-10,
xmax=10,
xtick={-8,-6,-4,-2,0,2,4,6,8},
xlabel style={font=\color{white!15!black}},
xlabel={\large Error [cm]},
ymin=0,
ymax=0.6,
ylabel style={font=\color{white!15!black}},
ylabel={\large PDF},
axis background/.style={fill=white},
legend style={legend cell align=left, align=left, draw=white!15!black}
]
\addplot[ybar interval, fill=mycolor1, fill opacity=0.6, draw=black, area legend] table[row sep=crcr] {%
x	y\\
-15	0\\
-14.5	0\\
-14	0\\
-13.5	0\\
-13	0\\
-12.5	0.00212992545260916\\
-12	0.00141995030173944\\
-11.5	0.00141995030173944\\
-11	0.00070997515086972\\
-10.5	0.00283990060347888\\
-10	0.0113596024139155\\
-9.5	0.00993965211217607\\
-9	0.0120695775647852\\
-8.5	0.00993965211217607\\
-8	0.0141995030173944\\
-7.5	0.0163294284700035\\
-7	0.0156194533191338\\
-6.5	0.0184593539226127\\
-6	0.022719204827831\\
-5.5	0.0262690805821796\\
-5	0.0241391551295705\\
-4.5	0.0326588569400071\\
-4	0.040468583599574\\
-3.5	0.0362087326943557\\
-3	0.0482783102591409\\
-2.5	0.0496982605608804\\
-2	0.063187788427405\\
-1.5	0.0823571175008875\\
-1	0.104366347177849\\
-0.5	0.134895278665247\\
0	0.163294284700035\\
0.5	0.182463613773518\\
1	0.207312744053958\\
1.5	0.200922967696131\\
2	0.144834930777423\\
2.5	0.0979765708200213\\
3	0.0575079872204473\\
3.5	0.0347887823926163\\
4	0.0212992545260916\\
4.5	0.0113596024139155\\
5	0.00780972665956692\\
5.5	0.00993965211217607\\
6	0.00922967696130635\\
6.5	0.00425985090521832\\
7	0.00567980120695776\\
7.5	0.00496982605608804\\
8	0.00212992545260916\\
8.5	0.00141995030173944\\
9	0.00567980120695776\\
9.5	0.00212992545260916\\
10	0.00141995030173944\\
10.5	0\\
11	0\\
11.5	0\\
12	0.00070997515086972\\
12.5	0\\
13	0\\
13.5	0\\
14	0\\
14.5	0\\
15	0\\
};
\addlegendentry{Error before enhancement}

\addplot[ybar interval, fill=mycolor2, fill opacity=0.6, draw=black, area legend] table[row sep=crcr] {%
x	y\\
-15	0\\
-14.5	0\\
-14	0\\
-13.5	0\\
-13	0\\
-12.5	0\\
-12	0\\
-11.5	0\\
-11	0\\
-10.5	0\\
-10	0\\
-9.5	0\\
-9	0\\
-8.5	0\\
-8	0\\
-7.5	0\\
-7	0\\
-6.5	0\\
-6	0\\
-5.5	0\\
-5	0\\
-4.5	0.00141995030173944\\
-4	0\\
-3.5	0.00070997515086972\\
-3	0.00425985090521832\\
-2.5	0.0170394036208733\\
-2	0.0440184593539226\\
-1.5	0.147674831380902\\
-1	0.230031948881789\\
-0.5	0.504082357117501\\
0	0.448704295349663\\
0.5	0.261270855520057\\
1	0.114305999290025\\
1.5	0.080937167199148\\
2	0.0468583599574015\\
2.5	0.0291089811856585\\
3	0.0333688320908768\\
3.5	0.0163294284700035\\
4	0.00496982605608804\\
4.5	0.00425985090521832\\
5	0.00425985090521832\\
5.5	0.00283990060347888\\
6	0.00141995030173944\\
6.5	0.00141995030173944\\
7	0.00070997515086972\\
7.5	0\\
8	0\\
8.5	0\\
9	0\\
9.5	0\\
10	0\\
10.5	0\\
11	0\\
11.5	0\\
12	0\\
12.5	0\\
13	0\\
13.5	0\\
14	0\\
14.5	0\\
15	0\\
};
\addlegendentry{Error after NN enhancement}
\end{axis}
\end{tikzpicture}%
}}
		\vspace{-1mm}
		\hspace{-2mm}\subfloat[Error distribution in stairs height estimation.\label{fig:deviations_height}]{\resizebox{\columnwidth}{!}{\begin{tikzpicture}

\begin{axis}[%
width=4.52in,
height=1.82in,
at={(0in,0in)},
scale only axis,
xmin=-10,
xmax=10,
xlabel style={font=\color{white!15!black}},
xlabel={\large Error [cm]},
ymin=0,
ymax=0.5,
ylabel style={font=\color{white!15!black}},
ylabel={\large PDF},
axis background/.style={fill=white},
legend style={legend cell align=left, align=left, draw=white!15!black}
]
\addplot[ybar interval, fill=mycolor1, fill opacity=0.6, draw=black, area legend] table[row sep=crcr] {%
x	y\\
-15	0\\
-14.5	0\\
-14	0\\
-13.5	0\\
-13	0\\
-12.5	0\\
-12	0.00496982605608804\\
-11.5	0.00496982605608804\\
-11	0.00496982605608804\\
-10.5	0.0070997515086972\\
-10	0.00780972665956692\\
-9.5	0.00212992545260916\\
-9	0.00496982605608804\\
-8.5	0.00993965211217607\\
-8	0.00425985090521832\\
-7.5	0.00993965211217607\\
-7	0.0120695775647852\\
-6.5	0.0312389066382677\\
-6	0.0383386581469649\\
-5.5	0.0525381611643592\\
-5	0.0610578629747959\\
-4.5	0.0653177138800142\\
-4	0.0773872914447994\\
-3.5	0.080937167199148\\
-3	0.0773872914447994\\
-2.5	0.0979765708200213\\
-2	0.080937167199148\\
-1.5	0.0681576144834931\\
-1	0.0830670926517572\\
-0.5	0.0795172168974086\\
0	0.0851970181043663\\
0.5	0.0681576144834931\\
1	0.0702875399361022\\
1.5	0.063187788427405\\
2	0.0674476393326234\\
2.5	0.0724174653887114\\
3	0.0468583599574015\\
3.5	0.0489882854100106\\
4	0.0624778132765353\\
4.5	0.0560880369187078\\
5	0.0425985090521832\\
5.5	0.0454384096556621\\
6	0.0468583599574015\\
6.5	0.0362087326943557\\
7	0.0383386581469649\\
7.5	0.040468583599574\\
8	0.0205892793752219\\
8.5	0.0291089811856585\\
9	0.0262690805821796\\
9.5	0.0234291799787007\\
10	0.0241391551295705\\
10.5	0.00496982605608804\\
11	0.00638977635782748\\
11.5	0.00496982605608804\\
12	0.00070997515086972\\
12.5	0.00141995030173944\\
13	0\\
13.5	0\\
14	0\\
14.5	0\\
15	0\\
};
\addlegendentry{Error before enhancement}

\addplot[ybar interval, fill=mycolor2, fill opacity=0.6, draw=black, area legend] table[row sep=crcr] {%
x	y\\
-15	0\\
-14.5	0\\
-14	0\\
-13.5	0\\
-13	0\\
-12.5	0\\
-12	0\\
-11.5	0\\
-11	0\\
-10.5	0\\
-10	0\\
-9.5	0\\
-9	0.00070997515086972\\
-8.5	0\\
-8	0\\
-7.5	0\\
-7	0.00070997515086972\\
-6.5	0.00070997515086972\\
-6	0.00212992545260916\\
-5.5	0.0035498757543486\\
-5	0.0134895278665247\\
-4.5	0.0312389066382677\\
-4	0.0553780617678381\\
-3.5	0.0695775647852325\\
-3	0.0575079872204473\\
-2.5	0.0830670926517572\\
-2	0.134895278665247\\
-1.5	0.2790202342918\\
-1	0.440894568690096\\
-0.5	0.327298544550941\\
0	0.198793042243521\\
0.5	0.157614483493078\\
1	0.0951366702165424\\
1.5	0.0383386581469649\\
2	0.0035498757543486\\
2.5	0.0035498757543486\\
3	0.00212992545260916\\
3.5	0\\
4	0.00070997515086972\\
4.5	0\\
5	0\\
5.5	0\\
6	0\\
6.5	0\\
7	0\\
7.5	0\\
8	0\\
8.5	0\\
9	0\\
9.5	0\\
10	0\\
10.5	0\\
11	0\\
11.5	0\\
12	0\\
12.5	0\\
13	0\\
13.5	0\\
14	0\\
14.5	0\\
15	0\\
};
\addlegendentry{Error after NN enhancement}
\end{axis}
\end{tikzpicture}%
}}
		\caption{The PDFs of the error between the estimated and real dimensions before and after NN enhancements. \label{fig:errors_dist}}
		\vspace{-5mm}
	\end{figure}

  Table.~\ref{t: init_error} and Fig.~\ref{fig:errors_dist} present the quantitative comparison using the average scores of the aforementioned metrics together with the probability density function (PDF) of the deviations in each dimension estimate. The NN-enhancement algorithm shows a clear improvement in overall evaluated scores by $\approx$ 75\%. This improvement is also reflected in the distributions where the NN-estimator shows a much better uncertainty than the initial dimensioning framework. These results support the previously stated arguments that the initial dimension estimation algorithm relies mainly on the captured radar raw features, which prompts the range and angular resolutions as the main controlling attributes for the estimation accuracy. Accordingly, the presented shallow NN can limit such hardware resolution obligations and model the error induced by the prosthetic leg motion. It follows that the proposed perception system can still achieve comparable accuracy declared by the state-of-the-art techniques ($\approx$ \unit[1]{cm}), regardless of the limitation in the MIMO angular resolution. 
	
	\begin{table}[t]
		\centering
		\caption{Quantitative analysis of the initial dimensioning vs NN-enhancement algorithm.}
		\label{t: init_error}
		\def\arraystretch{1.0}%1.5
		\begin{tabular}{lccc}
			\toprule
			& MAE / cm &  RMSE / cm & $\pm\sigma$ / cm\\
			\midrule
			Initial-Depth &2.78&4.77& 3.27\\
			Initial-Height&3.83&4.71&  4.50\\
			\bottomrule 
			\rule{0pt}{2ex}NN-Depth &0.81&1.21& 0.72\\
			NN-Height&1.19&1.59&0.75\\
			\bottomrule
		\end{tabular}
		\vspace{-5mm}
	\end{table}
	\vspace{-0.5mm}\section{Conclusion \& Future Work} \label{conc}
	\par In this study, a radar-based perception system is mounted to a prosthetic leg to develop a more natural stair-climbing activity. The main scope of the presented algorithm is estimating the stair height and depth while walking on a real-time basis. The dimensioning algorithm is independent of a specific gait cycle or specific knee positioning. The perception system is featured with a MIMO radar for scanning the sagittal plane and an IMU for coordinates correction due to the knee angular motion. The MIMO radar module is divided into 3 main microcontrollers, where the stair dimensioning algorithm is implemented. The dimensions are estimated by localizing two consecutive corners, which are used for estimating the depth and height. The estimated dimensions are enhanced by using a NN and there is no high-resolution radar hardware requirement. The final achieved output accuracy is revealing a MAE of \unit[0.81]{cm} for the depth and \unit[1.12]{cm} for the height. For future work, more scenarios will be taken into consideration, such as approaching the stairs in a lateral aspect. The presented algorithm will be extended to include the differentiation between more obstacles, e.g., ramps and curb-stones.
	
	\vspace{-0.5mm}
	\bibliographystyle{IEEEbib}
	
\end{document}